\documentclass{aastex}
\usepackage{emulateapj5}

\newcommand{\epcs}{{\rm erg\,cm^{-2}\,s^{-1}}}
\newcommand{\cts}{{\rm count\,s^{-1}}}
\newcommand{\src}{XTE~J0929$-$314}

\slugcomment{Accepted by ApJ Letters}

\shortauthors{Galloway et al.}
\shorttitle{XTE J0929$-$314}

\begin{document}

\title{Discovery of a High-Latitude Accreting Millisecond Pulsar
in an Ultracompact Binary}

\author{Duncan~K.~Galloway, Deepto~Chakrabarty\altaffilmark{1,2}, 
Edward~H.~Morgan, and Ronald~A.~Remillard}
\affil{\footnotesize Center for Space Research,
  Massachusetts Institute of Technology, Cambridge,MA 02139;}
\email{duncan,deepto,ehm,rr@space.mit.edu}

\altaffiltext{1}{Also Department of Physics, Massachusetts Institute of
Technology, Cambridge, MA 02139.}
\altaffiltext{2}{Alfred P. Sloan Research Fellow.}
\addtocounter{footnote}{-2}

\begin{abstract}
We have identified the third known accretion-powered millisecond
pulsar, XTE J0929$-$314, with the {\em Rossi X-Ray Timing Explorer}.
The source is a faint, high--Galactic-latitude X-ray transient 
($d\gtrsim 5$~kpc) that was in outburst during 2002
April--June.  The 185~Hz (5.4~ms) pulsation had a fractional rms
amplitude of 3--7\% and was generally broad and sinusoidal, although
occasionally double-peaked.  The hard X-ray pulses arrived up to
770~$\mu$s earlier than the soft X-ray pulses.  The pulsar was spinning
down at an average rate of $\dot\nu=(-9.2\pm 0.4)\times 10^{-14}$
Hz~s$^{-1}$; the spin-down torque may arise from magnetic coupling to
the accretion disk, a magnetohydrodynamic wind, or
gravitational radiation from the
rapidly spinning pulsar.  The pulsations were modulated by a 43.6~min
ultracompact binary orbit, yielding the smallest measured mass
function ($2.7\times 10^{-7} M_\odot$) of any stellar binary.  The
binary parameters imply a $\simeq 0.01 M_\odot$ white dwarf donor and
a moderately high inclination.  We note that all three known accreting
millisecond pulsars are X-ray transients in very close binaries with
extremely low mass transfer rates.  This is an important clue to the
physics governing whether or not persistent millisecond pulsations are
detected in low-mass X-ray binaries.
\end{abstract}


\section{INTRODUCTION}

Accretion-powered millisecond pulsars, the presumed progenitors of
millisecond radio pulsars, have proven surprisingly elusive for 20
years.  The first example, the X-ray transient SAX J1808.4$-$3658
($P_{\rm spin}=$401 Hz, $P_{\rm orb}=$2 hr), was identified as a
millisecond pulsar only four years ago \cite[]{wij98b,chak98d}.  This
only deepened the puzzle of why more examples are not known, since
SAX~J1808.4$-$3658 is very similar to many of the $\simeq50$ neutron
stars in low-mass X-ray binaries (LMXBs) that are not known pulsars
\cite[]{pc99}.  Earlier this year, a second system was detected,
the X-ray transient XTE J1751$-$305 ($P_{\rm spin}=$ 435 Hz, $P_{\rm
orb}=$42.4 min; Markwardt et al. 2002).  We report here on
the discovery of a third example, again with a very short binary
period.   

The faint X-ray transient XTE~J0929$-$314 ($l=260\fdg1$, $b=14\fdg2$)
was discovered by the All Sky Monitor (ASM) on the {\em Rossi X-Ray
Timing Explorer (RXTE)}\/ in 2002 April \cite[]{remillard02a}.  A
brief scanning observation with {\em RXTE}\/ detected persistent 185~Hz
pulsations \cite[]{remillard02b}, and further timing revealed a
circular, 43.6-min binary orbit \cite[]{gmrc02b}.  The high Galactic
latitude makes this source ideal for multiwavelength study.  Variable
optical \cite[]{gre02,cacella02} and radio \cite[]{rupen02}
counterparts were detected at the X-ray source position measured 
with the {\em Chandra X-Ray Observatory} (Juett et al.  2002, in
preparation), and \ion{C}{3}/\ion{N}{3} $\lambda\lambda$4640--4650 and
H$\alpha$ $\lambda$6563 emission lines were reported in the optical
spectrum \cite[]{cs02}.  In this Letter, we present a detailed
analysis of the {\em RXTE}\/ observations.   

\vspace*{0.2in}
\section{OBSERVATIONS}

XTE~J0929$-$314 is the faintest new transient discovered by the {\em
RXTE}\/ All-Sky Monitor \cite[ASM;][]{asm96}.  This 1.5--12~keV
instrument consists of three scanning shadow cameras which provide
90~s exposures of most points on the sky every 96~min.
XTE~J0929$-$314 was identified using a ``deep sky map'' technique, in 
which a map of flux residuals is constructed by cross-correlating the
predicted coded mask pattern against the best-fit data residuals for
each $4'$ cell in the field of view. Maps are made for each ASM camera,
using all of the low-background exposures in a given week.
This allows more
sensitive searches for new sources (as low as $\simeq 15$~mCrab away from
the Galactic center) compared to the $\simeq 50$ mCrab threshold for X-ray
error boxes derived from individual camera snapshots. 

After XTE~J0929$-$314 was identified by ASM, we obtained a series of
pointed {\em RXTE}\/ observations of the source during 2002 May 2 -- June
24 (MJD 52396--52449).  Our analysis is primarily based on data from the
{\em RXTE}\/ Proportional Counter Array (PCA; Jahoda et al. 1996), which
consists of five identical gas-filled proportional counter units (PCUs)
with a total effective area of $\approx6000\ {\rm cm}^2$ and sensitivity
to X-ray photons in the 2.5--60~keV range.  For all our PCA observations,
the data were collected in GoodXenon mode (in addition to the standard
data modes), which records the arrival time (1~$\mu$s
\centerline{\epsfxsize=8.5cm\epsfbox{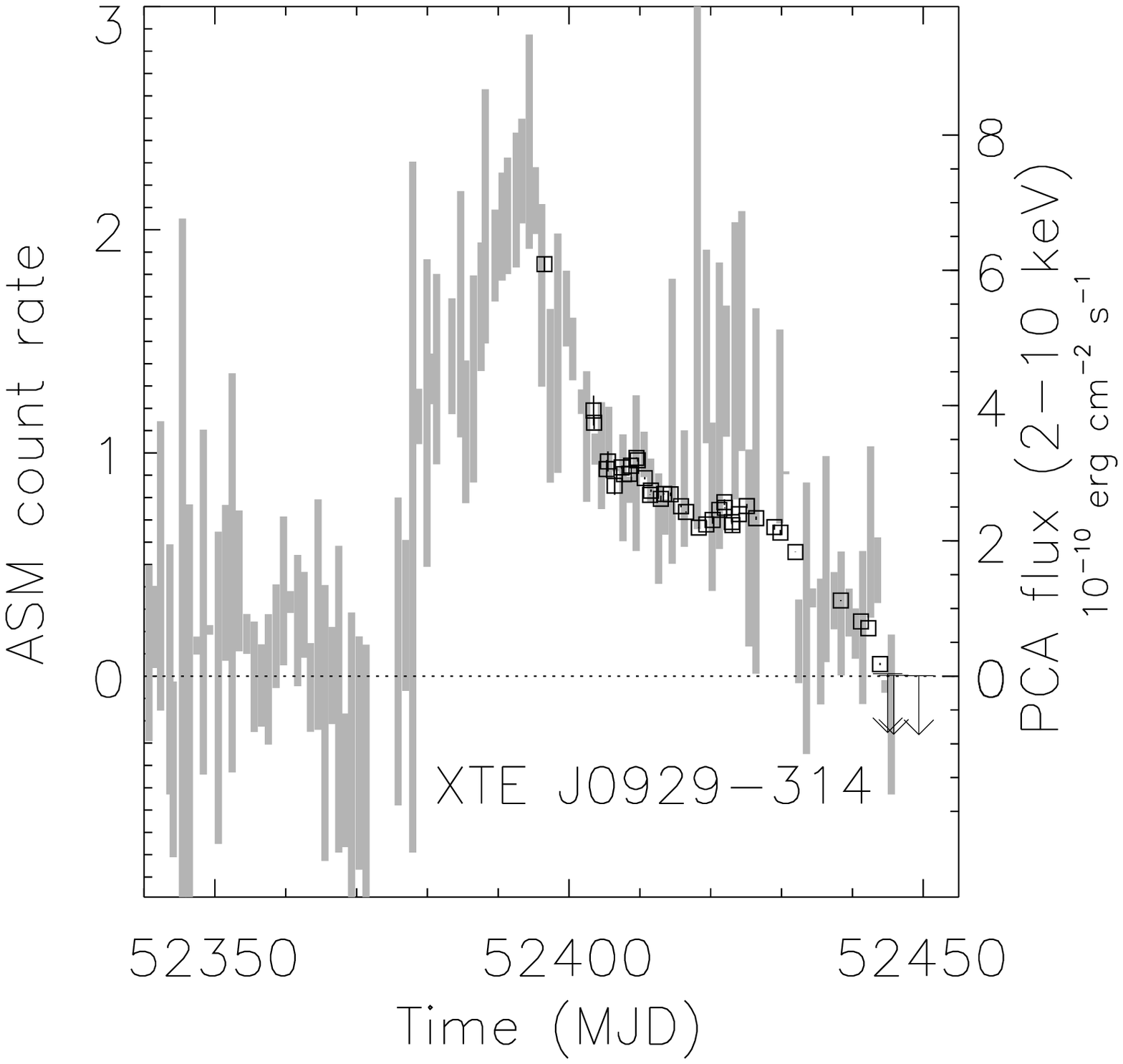}}
 \figcaption{2--10~keV ASM and PCA flux of \src\ throughout
the 2002 outburst.  The 1-d averaged ASM count rate $1\sigma$ confidence
intervals are shown as shaded regions (left $y$-axis), while the PCA
integrated flux is overplotted as open squares (right hand $y$-axes).
Estimated uncertainties on the PCA flux measurements are also shown, but are
typically smaller than the symbol size.
 \label{outburst} }
\bigskip
\noindent
resolution) and
energy (256 channel resolution) of every unrejected photon. We estimated
the background flux using
the ``CM'' faint-source model for PCA
gain epoch 5 (from 2000 May 13).  The photon arrival times at the
spacecraft were converted to barycentric dynamical times (TDB) at the
solar system barycenter using the Jet Propulsion Laboratory DE-200 solar
system ephemeris \cite[]{standish92} along with a spacecraft ephemeris and
{\em RXTE}\/ fine clock corrections. The position adopted was that of the
optical counterpart \cite[]{gre02}. Data from the 20--200~keV
High-Energy X-ray Timing Experiment \cite[HEXTE;][]{hexte96} were also
used to characterize the hard X-ray spectrum.

\section{ANALYSIS AND RESULTS}

The ASM and PCA intensity history of XTE J0929$-$314 is shown in Figure~1.
The ASM 1.5--12~keV count rate began rising around MJD~52370 and peaked at
$\approx 2.4~\cts$ ($\simeq$31~mCrab) on MJD~52394. The initial scanning
PCA observation occurred 2~d later, when the decay had already begun. Over
the following 20~d the source faded steadily, holding briefly at 10~mCrab
on MJD~52420 before declining below the detection limit. The last
3$\sigma$ PCA detection was on MJD~52443 at a flux of $1.8\times10^{-11}\
\epcs$ (2--10 keV).  For subsequent PCA observations, the flux was below
the $3\sigma$ detection threshold of $7.5\times10^{-12}\ \epcs$ (2--10~keV;
or $1.5\times10^{-11}\ \epcs$, 2--60~keV).
The estimated 2--60~keV fluence over the entire outburst was $4.2\times 10^{-3}\
{\rm erg\,cm^{-2}}$.  The combined PCA and HEXTE X-ray spectra were
consistent with an absorbed power law+blackbody model, with photon index
$\Gamma=$1.7--2.0, blackbody temperature $kT=$0.5--0.9~keV, and absorption
column density $N_{\rm H} < 10^{22}$~cm$^{-2}$. For most of the
observations we detected the source up to 50~keV. We estimated the
bolometric correction as the mean ratio of the integrated 2--60~keV to
2--10~keV flux, $2.34\pm0.12$.  No X-ray bursts were detected in the data. 

For our timing analysis, we selected photons from the top layer of each
PCU in the energy range 3--13~keV (absolute channels 7--30) in order
to maximize the pulsed signal to noise ratio.  We binned the arrival times
for these photons into $2^{-11}$~s ($\approx$0.5 ms) samples. An 185~Hz
pulsed signal was easily detectable in these data, with a fractional rms
amplitude of between 3 and 7\%.  The pulse profile in the initial PCA
observation on MJD 52396 contained a weak secondary maximum, but it was
broad and single-peaked in all subsequent observations except for the
interval MJD~52425--52430, when a second peak was again present (Figure
2).  The last 3$\sigma$ detection of pulsations was on MJD 52442, when the
source flux was $7.1\times10^{-11}\ \epcs$ (2--10 keV).  A $\simeq$1~Hz
quasi-periodic oscillation (FWHM=0.21~Hz) with $\simeq$5\% rms amplitude
was detected during most of the observations.

We investigated the energy-dependence of the pulse properties by
extracting pulse profiles in nine distinct energy bands from one of the
longer PCA observations on MJD 52408. The rms strength of the
pulsation decreased from around 5\% at 2--4~keV to 2.5\% at 10~keV.
Comparison of the pulse phase in each energy band indicated that
significant soft lags were present. For each energy band between 2.5
and 10.5~keV the pulse arrival time was earlier than in the adjacent
lower energy band, with a cumulative difference of 
770~$\mu$s.  The energy dependence of both the rms strength and the
phase lag of the pulsation appeared to flatten above 10~keV; there
were insufficient counts to explore the dependence above 20~keV.
Similar phase lags were observed in the millisecond pulsations from
SAX~J1808.4$-$3658, but that source did not show a similar energy
dependence for the pulsed amplitude (Cui, Morgan, \& Titarchuk 1998;
Ford 2000).  We will present a more detailed analysis of the pulse
shape variations elsewhere. 

\centerline{\epsfxsize=8.5cm\epsfbox{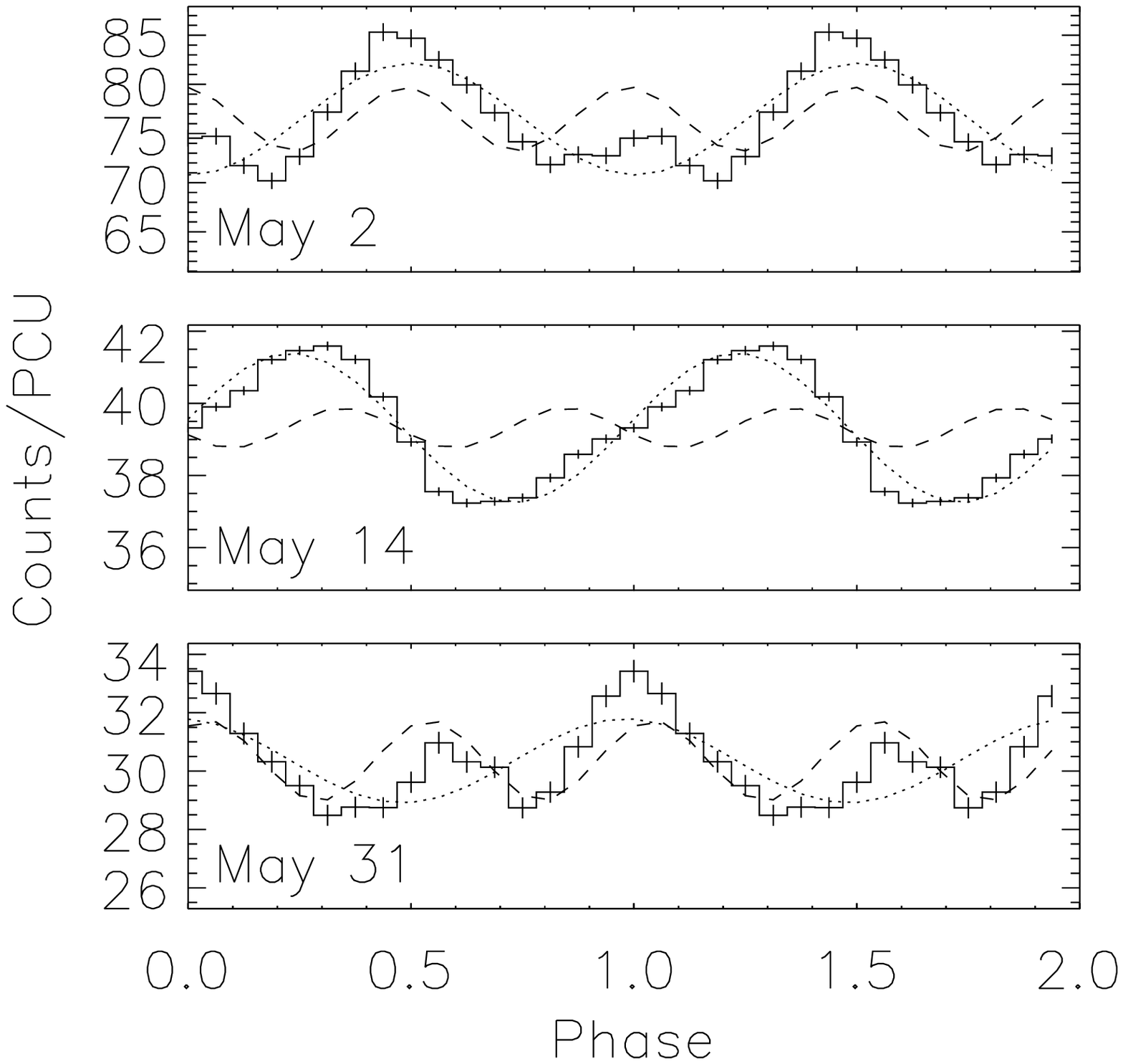}}
 \figcaption{Pulse profiles of \src\ measured on 2002 May 2,
14 and 31 (MJD 52396, 52408 and 52425). The profile is plotted as a solid
histogram, with error bars showing the $1\sigma$ uncertainties. The first
two Fourier components are overplotted as dotted lines (first harmonic)
and dashed lines (second harmonic). Note that the phase alignment between
the panels is arbitrary.
 \label{profile} }
\bigskip
\noindent

\centerline{\epsfxsize=8.5cm\epsfbox{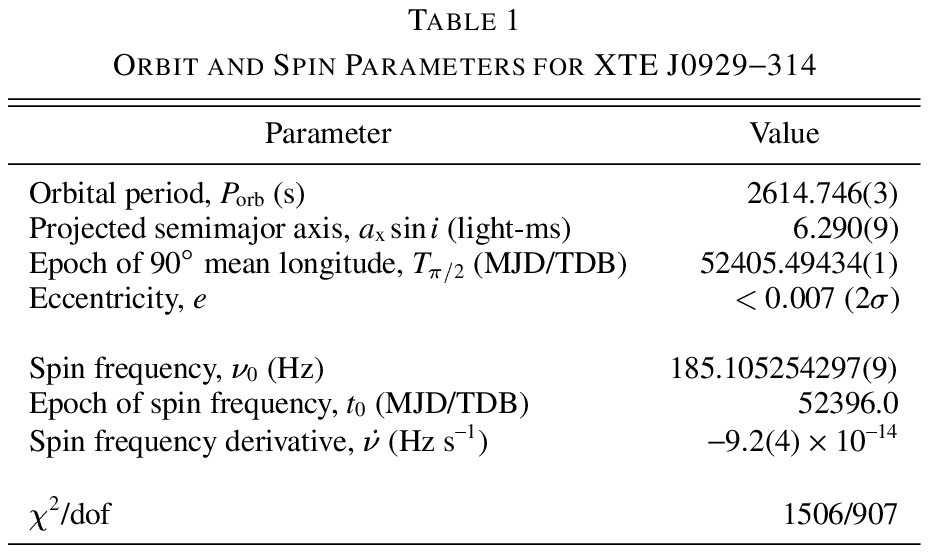}}
The 185~Hz pulsation showed a clear Doppler modulation, and we used a
Fourier frequency history to fit a preliminary orbital model (Galloway
et al. 2002).   We then applied a more precise analysis using the
pulse phase techniques commonly employed in radio pulsar timing (see,
e.g., Manchester \& Taylor 1977).  Adopting our 
preliminary orbital
model, we epoch-folded 43.6~min intervals of data (thus averaging over
periodic phase residuals caused by errors in our assumed orbital
parameters) and used the resulting linear drift in phase residuals to
refine our knowledge of the pulsar's spin frequency.  Then, we
combined this refined spin frequency and the preliminary orbital model
to epoch-fold 256~s intervals of data.  The resulting phase residuals
were used to compute differential corrections to our Keplerian orbit
model (see, e.g., Deeter, Boynton, \& Pravdo 1981). No significant
eccentricity was detectable, but a spin frequency derivative
substantially improved the fit.  Our best-fit orbit and spin parameters
for the pulsar are given in Table~1.  The pulse time delays due to the
orbit, as well as the effect of
fitting a constant-frequency model are shown in Figure~3.

\section{DISCUSSION}

From the presence of persistent millisecond pulsations over a wide
luminosity range in \src, we can infer an upper limit on the pulsar's
surface dipole magnetic
field strength of $B\lesssim 1\times
10^9\,d_{\rm 10kpc}$~G from accretion torque theory (Psaltis \&
Chakrabarty 1999).  This is consistent with the expectation that the
neutron star is a recycled 
pulsar whose magnetic field has decayed
during prolonged mass transfer (Bhattacharya \& van den Heuvel 1991).
It is interesting to 
note that the system has binary parameters that are extremely similar
to those of the other recently discovered millisecond X-ray pulsar XTE
J1751$-$305 (Markwardt et al. 2002) as well as the slow (7.6~s)
accreting X-ray pulsar 4U 1626$-$67 (Middleditch et al. 1981; 
Schulz et al. 2001), pointing to a similar evolutionary path.  A
puzzling aspect is that, unlike the two millisecond pulsars, the LMXB
4U~1626$-$67 has a strong ($3\times 10^{12}$ G) magnetic field in addition
to its slow spin period, which may indicate that its neutron star was
formed recently through accretion-induced collapse (see Yungelson,
Nelemans, \& van den Heuvel 2002 and references therein).

We can estimate a crude lower bound on the distance to XTE~J0929$-$314 by
first noting that the time-averaged mass transfer rate driven by
gravitational radiation in this 
\centerline{\epsfxsize=8.5cm\epsfbox{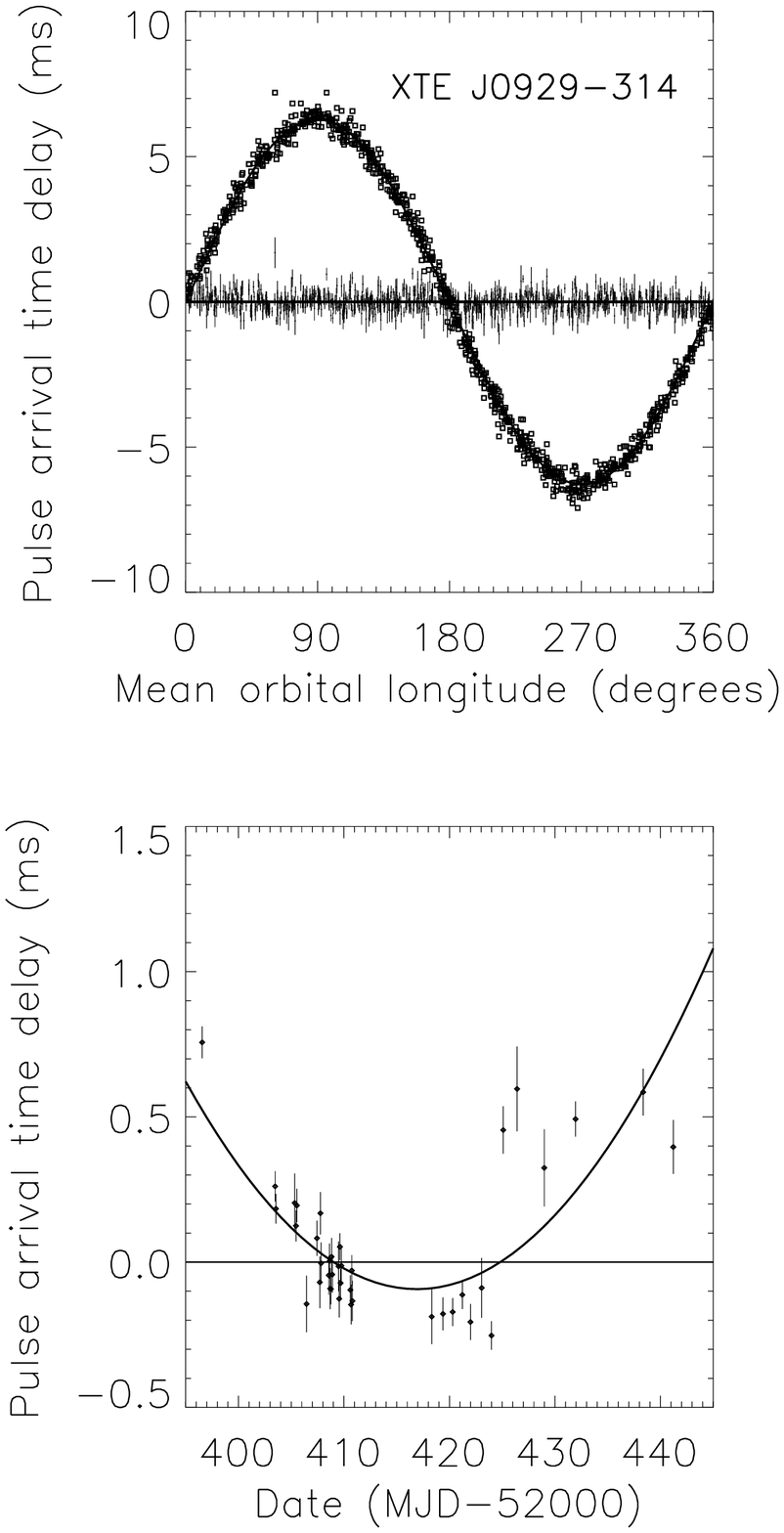}}
 \figcaption{Pulse timing residuals in \src. {\it Top
panel:} Fit residuals with and without the Keplerian orbit included. {\it
Bottom panel:} Fit residuals for a constant $\nu$ plus Keplerian orbit.
The best fit parabola to these residuals is also shown, indicating the
need for a $\dot{\nu}$ term in the model.  The especially large deviations
near MJD 52425 may be due to pulse shape variations.  Error bars represent
$1\sigma$ uncertainties.
 \label{resid} }
\bigskip
\noindent
43.6~min binary is (Faulkner 1971)
\begin{equation}
    \dot M_{\rm GW} = 5.5\times 10^{-12} 
          \left(\frac{M_{\rm x}}{1.4\,M_\odot}\right)^{2/3}
          \left(\frac{M_{\rm c}}{0.01\,M_\odot}\right)^2
\,M_\odot\mbox{\rm\,yr}^{-1} ,
\end{equation}
where $M_{\rm x}$ and $M_{\rm c}$ are the neutron star and companion
masses, respectively.  If we assume that the mass accretion rate during
the outburst is at least as large as $\dot M_{\rm GW}$, then our
non-detection at the end of the outburst requires a distance
$d\gtrsim 6$~kpc.
Also, given an outburst recurrence time of $\gtrsim 6.5$~yr (the time
since {\em RXTE}\/ was launched), the 2002 outburst fluence implies a
mean mass transfer rate equal to $\dot M_{\rm GW}$ for $d\gtrsim
5$~kpc.  We thus conclude that the source lies at least 1.2~kpc above the
Galactic plane.  Our limits indicate that the mass accretion rate $\dot M$
during the outburst was not more than a few percent of the Eddington
critical rate.  

The detection of spin-down during the outburst may provide an
opportunity to test accretion torque theory for X-ray pulsars in the
low-$B$ regime. The torque on a magnetic star from a prograde
accretion disk tends to spin-up the star for sufficiently high
$\dot M$, while accretion will be centrifugally inhibited (the
so-called ``propeller'' regime) for sufficiently low $\dot M$.
However, for an intermediate range of $\dot M$, the positive material
torque due to accretion may be dominated by a spin-down torque due to
one of a variety of mechanisms, even while accretion persists.
Two possible spin-down mechanisms include magnetic coupling of the
accretion disk and the magnetosphere (Ghosh \& Lamb 1979, 1991) and
expulsion of a centrifugally driven magnetohydrodynamic
wind (Anzer \& B\"orner 1980; Arons et al. 1984; Lovelace et
al. 1995). In either case, the torque should not exceed the
characteristic accretion torque $\dot M\sqrt{GM_{\rm x}r_{\rm co}}$
(where $r_{\rm co}$=50~km is the pulsar's corotation radius), which is
consistent with our measured $\dot\nu$ for $d\gtrsim 6$~kpc.  Another
possible mechanism is gravitational radiation by the rapidly spinning
pulsar (Bildsten 1998; Andersson, Kokkotas, \& Stergioulas 1999; Levin
1999).  Some of these issues can be explored by examining the relation
between $\dot M$ and $\dot\nu$, which has been done extensively for
high-$B$ neutron stars (e.g., Bildsten et al. 1997) but never for the
low-$B$ case.  A detailed analysis of the $\dot M$-$\dot\nu$
correlation will first require a better understanding of the pulse
shape variations, in order to limit systematic uncertainties in the
frequency history.  Such work is currently in progress.

White dwarfs and neutron stars accreting from a hydrogen-rich
companion cannot evolve to binary periods below about 80~min,
corresponding to the so-called ``period minimum'' observed for most
cataclysmic variables and LMXBs (Paczynski \& Sienkiewicz 1981;
Rappaport, Joss, \& Webbink 1982).  Ultracompact ($P_{\rm orb}\lesssim
80$~min) binaries like \src\ must therefore have a low-mass,
hydrogen-depleted (and probably degenerate) donor (Nelson, Rappaport,
\& Joss 1986).  Our measured orbital parameters further constrain the
nature of the donor in this system.   The pulsar mass function,
$f_{\rm X} =  2.7\times10^{-7}\ M_\odot$, is the smallest presently
known for any stellar binary.
This mass function gives a minimum companion mass ($i=90^\circ$) of
$M_{\rm c}=0.008$ $M_\odot$ for $M_{\rm x}=1.4\ M_\odot$; it also implies
$M_{\rm c}<0.03 M_\odot$ (95\% confidence) for a uniform a priori
distribution in $\cos i$ ($M_{\rm x}=2 M_\odot$).  The mass-radius
relation for a Roche-lobe--filling donor in a 43.6~min binary is $R_{\rm
c} = 0.04 (M_{\rm c}/0.01\, M_\odot)^{1/3}\,R_\odot$ \cite[]{ffw72}.  As
expected, this has no intersection with the theoretical mass-radius
relation for very low-mass hydrogen-rich stars (i.e. brown dwarfs,
Chabrier et al. 2000; cf. Bildsten \& Chakrabarty 2001). A very
low-mass cold helium white dwarf (Zapolsky \& Salpeter 1969; Nelemans et
al. 2001) is easily consistent for masses of $\simeq 0.01\,M_\odot$,
implying that the system has a relatively high (and thus probable)
inclination. We note that a cold carbon white dwarf is not consistent with
our measured orbital parameters (Lai, Abrahams, \& Shapiro 1991).  It may
be possible for a helium dwarf donor to retain a small residual hydrogen
content (Podsiadlowski, Rappaport, \& Pfahl 2002); this is of particular
interest, given the report of an H$\alpha$ $\lambda$6563 emission line in
the optical spectrum (Castro-Tirado et al. 2002).

Binary evolution theory predicts that many (if not most) of the
$\simeq$50 known neutron stars in LMXBs are spinning at millisecond 
periods (Bhattacharya \& van den Heuvel 1991).  Of these, only three are
now known pulsars, with persistent millisecond pulsations during their
X-ray active states.\footnote{ Another ten NS/LMXBs show millisecond
oscillations during thermonuclear X-ray bursts; these are probably also
signatures of rotation (see, e.g., Strohmayer \& Markwardt 2002).}
All three are low-luminosity transients in very close binaries, with
extremely small time-averaged $\dot M$; evidently, millisecond
pulsations are easier to detect in such systems.   This supports the
suggestion that magnetic screening by freshly accreted material may
prevent the formation of persistent X-ray pulses in NS/LMXBs with
$\dot M$ above a critical value (Cumming, Zweibel, \& Bildsten 2001). 

\acknowledgments We are grateful to Jean Swank, Evan Smith, and the {\em
RXTE}\/ operations team at NASA/GSFC for their help in scheduling these
target-of-opportunity observations.  We also thank Lars Bildsten, Fred
Lamb, Al Levine, Dimitrios Psaltis and Saul Rappaport for useful
discussions.  This work was supported in part by NASA under grant NAG
5-9184 and contract NAS 5-30612.

\end{document}